\documentclass[twocolumn]{aastex62}
\usepackage{amsmath}
\usepackage{amsbsy}

\hypersetup{linkcolor=magenta,citecolor=blue,filecolor=cyan,urlcolor=purple}

\received{2019 April 9}
\revised{2019 June 3}
\accepted{2019 June 7}

\shorttitle{Modeling a Stellar Superflare from $\kappa^{1}Cet$}
\shortauthors{Lynch et al.}

\begin{document}


\title{Modeling a Carrington-scale Stellar Superflare and Coronal Mass Ejection from $\kappa^{1}Cet$}


\correspondingauthor{Benjamin~J.~Lynch}
\email{blynch@ssl.berkeley.edu}

\author[0000-0001-6886-855X]{Benjamin~J.~Lynch}
\affiliation{Space Sciences Laboratory, University of California--Berkeley, Berkeley, CA 94720, USA}

\author[0000-0003-4452-0588]{Vladimir~S.~Airapetian}
\affiliation{NASA Goddard Space Flight Center, Greenbelt, MD 20771, USA}      
\affiliation{Department of Physics, American University, Washington, D.C. 20016, USA}

\author[0000-0002-4668-591X]{C.~Richard~DeVore}
\affiliation{NASA Goddard Space Flight Center, Greenbelt, MD 20771, USA}   

\author[0000-0001-8975-7605]{Maria~D.~Kazachenko}
\affiliation{Laboratory for Atmospheric and Space Physics, University of Colorado, Boulder, CO 80303, USA}

\author{Teresa~L\"{u}ftinger}
\affiliation{Department of Astrophysics, University of Vienna, Vienna, Austria}

\author[0000-0003-3061-4591]{Oleg~Kochukhov}
\affiliation{Department of Physics and Astronomy, Uppsala University, Uppsala, Sweden}

\author{Lisa~Ros\'{e}n}
\affiliation{Department of Physics and Astronomy, Uppsala University, Uppsala, Sweden}

\author{William~P.~Abbett}
\affiliation{Space Sciences Laboratory, University of California--Berkeley, Berkeley, CA 94720, USA}

\begin{abstract}

Observations from the \emph{Kepler} mission have revealed frequent superflares on young and active solar-like stars. Superflares result from the large-scale restructuring of stellar magnetic fields, and are associated with the eruption of coronal material (a coronal mass ejection, or CME) and energy release that can be orders of magnitude greater than those observed in the largest solar flares. These catastrophic events, if frequent, can significantly impact the potential habitability of terrestrial exoplanets through atmospheric erosion or intense radiation exposure at the surface.  We present results from numerical modeling designed to understand how an eruptive superflare from a young solar-type star, $\kappa^{1}Cet$, could occur and would impact its astrospheric environment. Our data-inspired, three-dimensional magnetohydrodynamic modeling shows that global-scale shear concentrated near the radial-field polarity inversion line can energize the closed-field stellar corona sufficiently to power a global, eruptive superflare that releases approximately the same energy as the extreme 1859 Carrington event from the Sun. We examine proxy measures of synthetic emission during the flare and estimate the observational signatures of our CME-driven shock, both of which could have extreme space-weather impacts on the habitability of any Earth-like exoplanets. We also {speculate} that the observed 1986 Robinson-Bopp superflare from $\kappa^{1}Cet$ {was perhaps} as extreme for that star as the Carrington flare was for the Sun.

\end{abstract}


\section{Introduction}

There is growing appreciation that planetary atmospheric chemistry, and even the retention of an atmosphere in many cases, depends critically upon the high-energy radiation and particle environments around the host star \citep{Segura2005, CohenO2014, DomagalGoldman2014, Rugheimer2015}. This has led to a number of increasingly sophisticated modeling efforts to characterize the space environment of exoplanets and stellar-wind/exoplanet interactions \citep{CohenO2011, Vidotto2011, Vidotto2012,  Garraffo2016, Garraffo2017, GarciaSage2017}. The cumulative effects of both steady-state and extreme space weather from active stars, including intense X-ray and extreme ultraviolet (EUV) radiation, large fluxes of highly energetic particles, and frequent exoplanet interactions with stellar coronal mass ejections (CMEs), will have a significant impact on the exoplanets' atmospheric evolution, and ultimately on their habitability \citep{Lammer2007, Lammer2009, DrakeJJ2013, Osten2015, Kay2016, Airapetian2019}. Additionally, stellar activity from young solar-type stars, such as $\kappa^{1} Cet$, is now being considered in investigations of the evolution of our own solar system, where enhanced levels of extreme space weather may have profoundly affected the chemistry and climate of the early Earth \citep{Airapetian2016a, Airapetian2017a, DongC2017a, GarciaSage2017}.

Solar flares---the explosive release of energy in the solar atmosphere across a wide range of electromagnetic wavelengths---occur due to the rapid release of free energy stored in the strong sheared and/or twisted magnetic fields typically associated with sunspots and active regions \citep{Forbes2000a, Fletcher2011, Kazachenko2012}. Solar flares and CMEs are widely, although not universally, accepted as being driven by magnetic reconnection \citep{Klimchuk2001,Lynch2008,Karpen2012}. The long-standing CSHKP model \citep{Carmichael1964, Sturrock1966, Hirayama1974, Kopp1976} for eruptive solar flares explains many of their generic observed properties \citep[e.g.][and references therein]{Janvier2015,Lynch2016b,Welsch2018}.

The total energy released during solar flares typically ranges over $E \sim 10^{29-32}$~erg \citep{Emslie2012}, whereas stellar flares can extend to a much higher range of energies $E \sim 10^{32-36}$~erg \citep{Shibata2013, Maehara2015, Notsu2019}. \emph{Kepler} observations of superflaring solar-type stars indicate that large starspots typically are associated with their flares, the frequency and maximum energy of the flares depend critically upon the age of the star, and younger stars exhibit greater maximum flare energies and higher flare frequencies \citep{Maehara2012, Notsu2013, Notsu2019, Shibayama2013}.

\citet{Schaefer2000} presented some of the earliest superflare observations associated with solar-type stars, including $\kappa^{1} Cet$, a G5 young solar analog aged $\sim 0.7$~Gyr. They estimated the $\kappa^{1} Cet$ superflare energy as $E \sim 2 \times 10^{34}$~erg, based on observations by \citet{Robinson1987} of \ion{He}{1} emission. $\kappa^{1} Cet$ is reported as having magnetic cycles and magnetic field strengths in the kiloGauss range \citep{Saar1992} and showing evidence of starspots that rotate differentially across the stellar disk \citep{Rucinski2004, Walker2007}. The star's quasi-steady wind and radiation environments are objects of ongoing study for their effects on planetary habitability \citep[e.g.][]{Ribas2010, doNascimento2016}.

In this paper, we present results from a three-dimensional numerical magnetohydrodynamics (MHD) simulation of a global stellar superflare and its CME, based on observations by \citet{Rosen2016} of the magnetic field of $\kappa^1 Cet$. Although stellar active-region fields associated with large starspots are thought to contain the magnetic free energy necessary to power superflares and their associated eruptions, stellar magnetogram observations tend to leave such strong-field regions unresolved. Our approach is to model self-consistently the gradual accumulation of free magnetic energy via the introduction of large-scale stresses to the observed global stellar field. In this way, we maximize both the amount of free energy available for release during the eruptive flare and the spatial scale of the stellar CME, without making any assumptions about the existence or strength of unresolved starspots on $\kappa^1 Cet$. Therefore, our global-scale eruption represents the most extreme stellar space-weather event possible within the observed constraints on the surface magnetic flux distribution, and sets a baseline for comparison with observations of superflares on $\kappa^1 Cet$ and similar stars.

Our paper is organized as follows. In section~\ref{sec:arms} we present the MHD numerical model. Section~\ref{sec:pre} describes the pre-eruption phase of the simulation, including the magnetic field configuration ($\S$\ref{sec:pre:mag}), the stellar-wind outflow ($\S$\ref{sec:pre:sw}), and the self-consistent, slowly driven energy-accumulation phase ($\S$\ref{sec:pre:shear}). In section~\ref{sec:cme} we present the eruption results and their analysis: the global-eruption overview and energy evolution ($\S$\ref{sec:cme:energy}), the stellar-flare reconnected flux ($\S$\ref{sec:cme:rxn}), the synthetic X-ray and EUV emission proxies ($\S$\ref{sec:cme:flare}), and the properties of the CME-driven shock ($\S$\ref{sec:cme:shock}). Section~\ref{sec:disc} concludes with a discussion of our results in the context of future modeling and observations of stellar space weather.


\section{Adaptively Refined MHD Solver (ARMS)}
\label{sec:arms}

The ARMS code, developed by \citep{DeVore2008} and collaborators, calculates solutions to the 3D nonlinear, time-dependent MHD equations that describe the evolution and transport of density, momentum, energy, and magnetic flux throughout the system. The numerical scheme used is a finite-volume, multidimensional flux-corrected transport algorithm \citep{DeVore1991}. ARMS is fully integrated with the adaptive-mesh toolkit PARAMESH \citep{MacNeice2000} to handle dynamic, solution-adaptive grid refinement and enable efficient multiprocessor parallelization.

For our simulation, we use ARMS to solve the following ideal MHD equations in spherical coordinates
\begin{equation}
    \frac{\partial \rho}{\partial t} + \nabla \cdot \left( \rho
    \boldsymbol{V} \right) = 0 , 
    \label{eq1}
\end{equation}
\begin{equation}    
    \frac{\partial}{\partial t} \left( \rho \boldsymbol{V} \right)
    + \nabla \cdot \left( \rho \boldsymbol{V} \boldsymbol{V} \right)
    + \nabla P = \frac{1}{4\pi}\left( \nabla \times \boldsymbol{B}
    \right) \times \boldsymbol{B} - \rho \boldsymbol{g} , 
    \label{eq2}
\end{equation}
\begin{equation}
    \frac{\partial \boldsymbol{B}}{\partial t} = \nabla \times
    \left( \boldsymbol{V} \times \boldsymbol{B} \right), 
    \label{eq4}
\end{equation}
where all the variables retain their usual meaning, solar gravity is $\boldsymbol{g} = g_\odot (r/R_\odot)^{-2} \boldsymbol{\hat{r}}$ with $g_\odot = 2.75\times10^4$~cm~s$^{-2}$, and we use the ideal gas law $P=2(\rho/m_p)k_BT$.  Given the isothermal approximation used in the construction of our background solar wind, we do not solve an internal energy or temperature equation. The plasma temperature remains uniform throughout the domain for the duration of the simulation.

Additionally, while there is no explicit magnetic resistivity in the equations of ideal MHD, necessary, and stabilizing numerical diffusion terms introduce an effective resistivity on very small spatial scales, i.e., the size of the grid. In this way, magnetic reconnection can occur when current sheet features and the associated gradients of field reversals have been distorted and compressed to the local grid scale.

The spherical computational domain uses logarithmic grid spacing in $r$ and uniform grid spacing in $\theta, \phi$. The domain extends from $r \in \left[1R_\odot, 30R_\odot\right]$, $\theta \in \left[ 11.25^{\circ}, 168.75^{\circ} \right]$ ($\pm 78.75^{\circ}$ in latitude), and $\phi \in \left[-180^{\circ}, +180^{\circ} \right]$ (longitude).  The initial grid consists of $6 \times 6 \times 12$ blocks with 8$^3$ grid cells per block. There are two additional levels of static grid refinement. The highest refinement region (level 3) is $r \in \left[1R_\odot, 5.485R_\odot\right]$ for all $\theta, \phi$, and the level 2 refinement extends from $r \in \left[5.485R_\odot, 9.650R_\odot\right]$. The maximum resolution is therefore $192 \times 192 \times 384$ with the level 3 grid cells having an angular width of $0.820^{\circ} \times 0.938^{\circ}$ in $\theta, \phi$ and a radial extent of $\Delta r = 0.01787R_\odot$ at the lower boundary. After the solar wind relaxation phase we turn on the adaptive-mesh refinement criteria. The maximum refinement remains at level 3 but as the eruption progresses and regions of high electric current density evolve, the computational grid refines and de-refines to track the evolution of the strong currents. The refinement criteria are described in \citet{Karpen2012}.

\section{Pre-Eruption Stellar Corona}
\label{sec:pre}

\subsection{Magnetic Field Configuration}
\label{sec:pre:mag}

For decades, the fossil magnetic fields of massive, early-type stars were analyzed assuming simple dipole or dipole-plus-quadrupole magnetic field geometries. For late-type active stars, the development of Zeeman-Doppler imaging \citep[ZDI;][]{Dontai1997,Piskunov2002} has made it possible to invert time series of high-resolution spectroscopic and spectropolarimetric Stokes observations (circularly and linearly polarized stellar light) into surface maps of parameters such as brightness, temperature, elemental abundance, and complex magnetic field geometry \citep{Lueftinger2010b,Lueftinger2010a}. ZDI has further matured and can now be used to determine the strength, distribution, polarity, and polarity reversals of surface magnetic fields \citep{Lueftinger2015,Kochukhov2016}. The availability of stellar magnetic field maps has significantly advanced our capacity for sophisticated numerical modeling of stellar coronae, winds, and star-planet interactions \citep{CohenO2011,Vidotto2011,doNascimento2016,Garraffo2016}.

\begin{figure}
\centering
\includegraphics[width=0.47\textwidth]{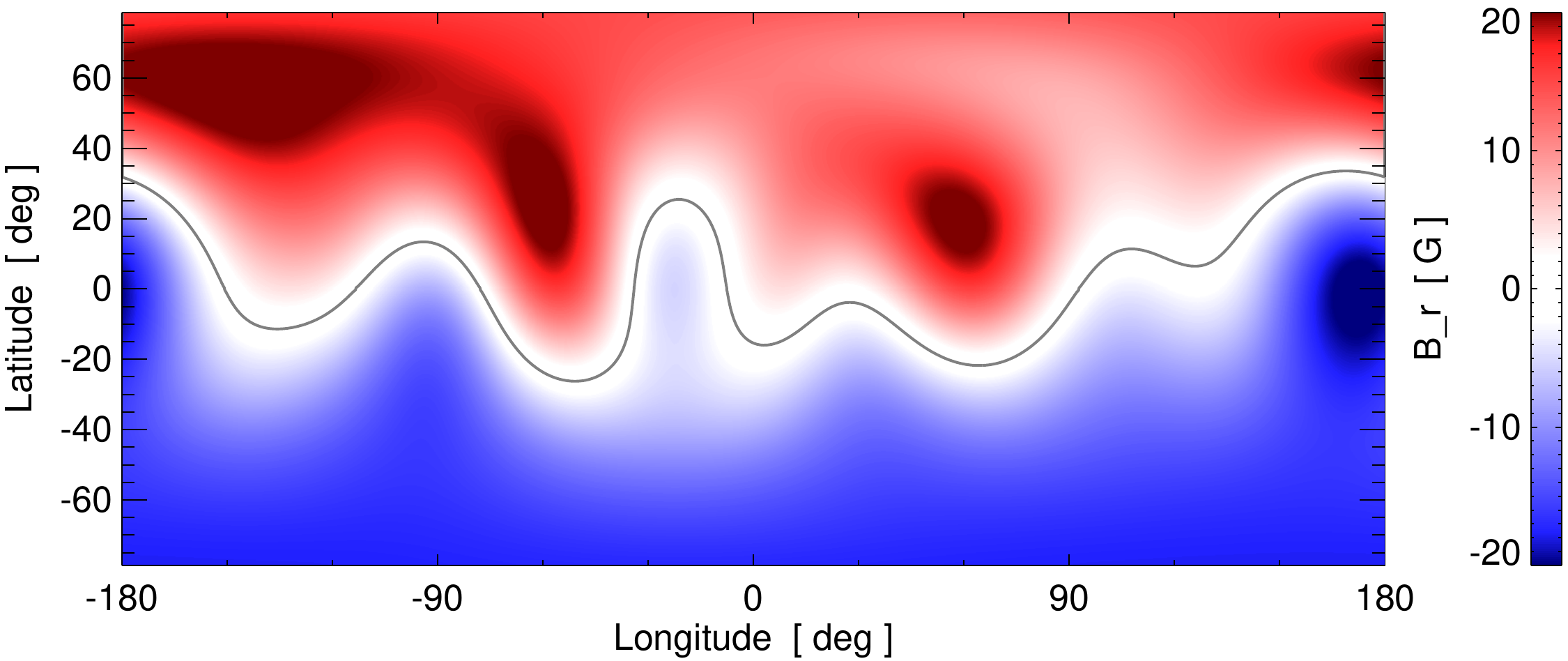}
\caption{$B_r(\theta,\phi)$ distribution derived from 2012 August Zeeman-Doppler imaging observations of $\kappa^1Cet$ \citep{Rosen2016} used to calculate the simulation's initial magnetic configuration. 
}
\label{f1}
\end{figure}

In our numerical simulation, we use the $\kappa^{1} Cet$ surface field distribution from \citet{Rosen2016} during the epoch 2012.9 (mid-to-late August) derived from the ZDI analysis of data from PolarBase \citep{Petit2014}. Figure~\ref{f1} shows this $B_r(\theta,\phi)$ distribution at the lower radial boundary. Positive radial field is denoted as red, negative as blue, and the black line indicates the $B_r=0$ polarity inversion line (PIL). We generate the potential field source surface \citep[PFSS;][]{WangYM1992} solution which is used to initialize the magnetic field throughout the computational domain.

\subsection{Stellar Wind Outflow}
\label{sec:pre:sw}

The stellar properties of $\kappa^{1} Cet$ are discussed by \citet{doNascimento2016} in a study where they present results from a data-driven MHD simulation utilizing a polytropic ($\gamma=1.1$) stellar-wind model \citep{Vidotto2012} to show a mass-loss rate about 50 times greater that that of the Sun. While more advanced stellar-wind treatments are under development \citep[e.g.][]{Airapetian2019b}, our focus is on the storage and release of magnetic energy in the stellar corona during the eruptive superflare. Since our background stellar wind needs only to create the distinct open- and closed-flux systems characteristic of solar and stellar coronae, here we use the isothermal \citet{Parker1958} wind solution for a uniform temperature, $T_0 = 2\times10^6$~K.


The solar wind is initialized in ARMS by first solving the one-dimensional \citet{Parker1958} equation for a spherically symmetric isothermal corona,
\begin{equation}
    \frac{V_{\rm sw}^2}{c_0^2} - \ln\left( \frac{V_{\rm sw}^2}{c_0^2}
    \right) = -3 + 4 \ln\left( \frac{r}{r_c} \right) + 4 \frac{r_c}{r},
\end{equation}
where the base number density, pressure, and temperature are 
$n_0 = \rho_0/m_p = 9.05 \times 10^8$~cm$^{-3}$, 
$P_0 =0.5$~dyn~cm$^{-2}$, and 
$T_0 = 2.0\times10^6$~K, respectively.
Here, 
$c_0 = (2 k_B T_0 / m_p)^{1/2} = 181.7$~km~s$^{-1}$ 
is the thermal velocity at $T_0$ and the location of the critical point is
$r_c = G M_\odot / 2 c_0^2 = 2.87R_\odot$.  
These parameters yield a solar wind speed at the outer boundary of 
$V_{\rm sw}(30R_\odot) \approx 550$~km~s$^{-1}$.

%
\begin{figure}
\centering
\includegraphics[width=0.47\textwidth]{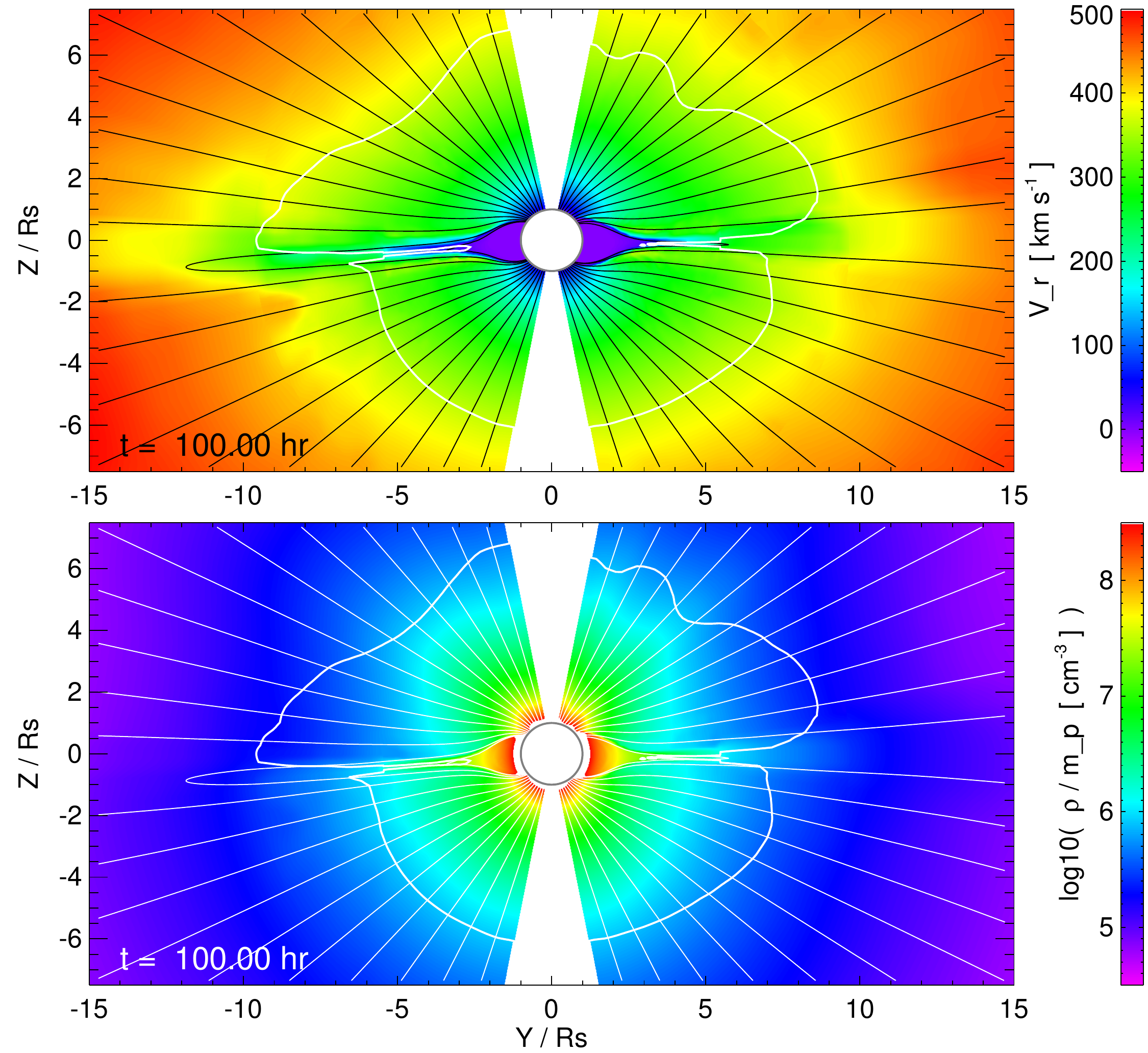}
\caption{Radial velocity (top) and plasma number density (bottom) at the end of the relaxation phase ($t_{\rm rel} = 100$~hr). Plane of the sky viewpoint is from the $0^{\circ}$ longitude central meridian. Representative field lines are shown to illustrate the structure and evolution of the open-field regions as the model stellar-wind outflow reaches a quasi-steady state.}
\label{fs1}
\end{figure}

At time $t=0$~s we impose this Parker $V_{\rm sw}(r)$ profile and use it to set the initial mass density profile $\rho(r)$ from the steady mass-flux condition ($\rho V_{\rm sw} r^2 =$~constant) throughout the computational domain. We then let the system relax to time $t_{\rm rel} = 3.6\times10^5$~s (100~hr). The initial discontinuities in the PFSS magnetic field solution at the source surface ($r=2.5R_\odot$) propagate outwards and eventually through the outer boundary.  During the relaxation phase, the highest layers of the closed streamer belt flux are carried outward by the stellar wind flow, setting up the condition for the transverse pressure from the open fields to push in behind the expanding streamer belt structure forming the elongated current sheet.  Eventually, the numerical diffusion allows magnetic reconnection between the elongated streamer belt field lines and gives the system the opportunity to adjust the amount of open flux relative to the new pressure balance associated with the background stellar wind outflow.  The inner-boundary mass source allows material to accumulate in the closed-field regions and sets up steady radial outflow along open field lines. Figure~\ref{fs1} shows a snapshot of the radial velocity (top panel) and the logarithmic number density (bottom panel) at the end of the relaxation phase ($t_{\rm rel} = 100$~hr). The Alfv\'{e}n surface is shown as a white contour and representative magnetic field lines are shown in each plot.

%
\begin{figure}
\centering
\includegraphics[width=0.47\textwidth]{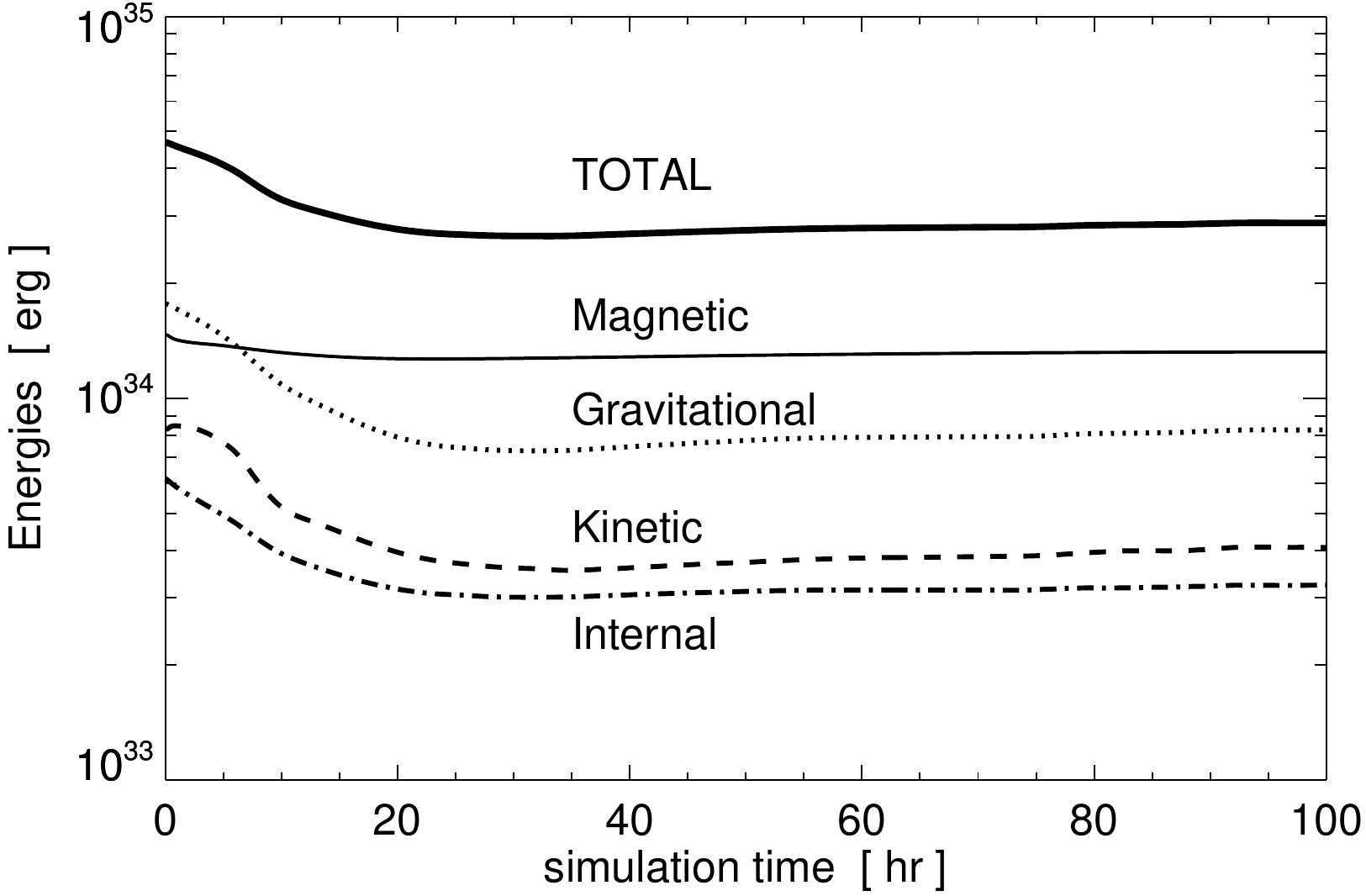}
\caption{Global energy evolution during solar wind relaxation phase ($0 \le t \le 100$~hr).}
\label{fs2}
\end{figure}

Figure~\ref{fs2} plots the global energy evolution during the relaxation phase: internal $E_{\rm int}$ (dash-dotted line), kinetic $E_{\rm K}$ (dashed line), gravitational $E_{\rm grv}$ (dotted line), magnetic $E_{\rm M}$ (solid thin line), and total $E_{\rm total}$ (solid thick line). The global energy curves reflect the initial PFSS magnetic configuration and the isothermal Parker wind outflow equilibrating, as described above. By the end of the solar wind relaxation phase, the global energies are
$E_{\rm int} = 3.24 \times 10^{33}$~erg, 
$E_{K} = 4.07 \times 10^{33}$~erg, 
$E_{\rm grv} = 8.26 \times 10^{33}$~erg,
$E_{M} = 1.32 \times 10^{34}$~erg, 
and 
$E_{\rm total} = 2.88 \times 10^{34}$~erg.
We can estimate a globally-averaged plasma beta as $\langle \beta \rangle = E_{\rm int}/E_M = 0.245$. We note, however, that the plasma $\beta$ is on the order of $10^{-3}$ in the strong-field regions of our stellar corona. Similarly, a globally averaged Alfv\'{e}n speed can be calculated as $\langle {V_A} \rangle = \sqrt{2E_M/M_{\rm tot}}$ where $M_{\rm tot}$ is the total mass of the system. Once the mass density profiles along open-field flux tubes reach quasi-equilibrium at the end of the relaxation phase, $M_{\rm tot}(t_{\rm rel}) = 6.548 \times 10^{18}$~g, yielding $\langle V_A \rangle = 635$~km~s$^{-1}$. While the stellar wind outflow is both supersonic and super-Alfv\'{e}nic for $r \gtrsim 5-10R_\odot$ (Figure~\ref{fs1}), the globally averaged Alfv\'{e}n speed can be compared to the CME and CME-driven shock velocities showing that our entire global CME eruption and its evolution proceeds at or exceeds $\langle V_A \rangle$.

\subsection{Accumulation of Magnetic Free Energy}
\label{sec:pre:shear}

Idealized surface flows are imposed on the lower radial boundary of our simulation to accumulate magnetic free energy in our stellar corona system. The boundary flows are constructed to follow the contours of $B_r(\theta,\phi)$ exactly so that the stellar radial flux distribution remains constant throughout the simulation as in \citet{DeVore2008}. This ensures that the potential magnetic energy of our system remains constant during the simulation and therefore any increase in the magnetic energy represents free magnetic energy that will be available for the stellar eruption. While these flows are obviously simpler than the complex photospheric motions observed on our Sun \citep{Li2004}, the cumulative effect of both the largest-scale photospheric motions of differential rotation and meridional flows \citep{vanBallegooijen1998,Sheeley2005,Yeates2014} acting on emerged active-region fields and the smallest-scale granulation, diffusion, and chaotic rotational motions of helicity condensation \citep{Antiochos2013,Knizhnik2017} is to form elongated, global-scale PILs with highly concentrated, non-potential, sheared field structures.

%
\begin{figure}
\centering
\includegraphics[width=0.47\textwidth]{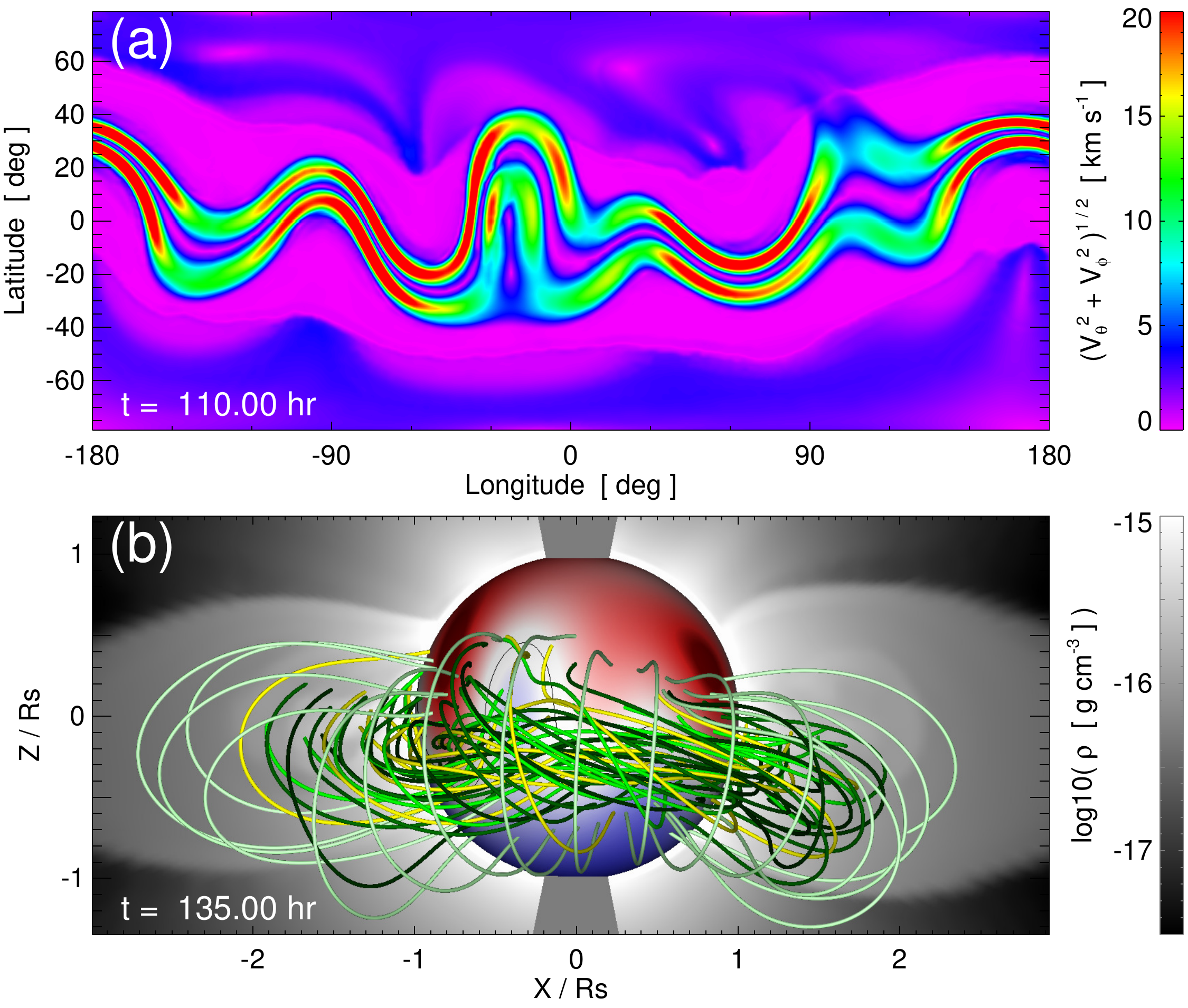}
\caption{(a) Tangential velocity magnitude $(V_\theta^2 + V_\phi^2)^{1/2}$ at the lower radial boundary at $t=110$~hr showing the energization phase shearing flow pattern $|\boldsymbol{V}_{\rm shear}|$ near the global polarity inversion line. (b) Representative magnetic field lines at$t=135$~hr showing the global-scale energized field structure.}
\label{fsFLOW}
\end{figure}

The boundary shearing flow profiles are as given as
\begin{equation}
    \boldsymbol{V}_{\rm shear}(\theta,\phi,t) = V_0 
    f_T f_B \left [ \boldsymbol{\hat{r}} \times \boldsymbol{\nabla_t} B_r(\theta,\phi) \right ] ,
\end{equation}
where $\boldsymbol{ \nabla_t }$ is the tangential gradient operator,
\begin{equation}
\boldsymbol{\nabla_t} = \frac{1}{R_\odot}\frac{\partial}{\partial \theta} \boldsymbol{\hat{\theta}} + \frac{1}{R_\odot \sin \theta}\frac{\partial}{\partial \phi} \boldsymbol{\hat{\phi}} ,
\end{equation} 
acting on $B_r$ at the $r=R_\odot$ boundary. The temporal dependence is given by
\begin{equation}
    f_T(t) = \left \{ \begin{array}{ll}
    		\frac{1}{2}-\frac{1}{2}\cos{\left[ \pi \frac{(t-100)}{5} \right]} \;\; & {\rm for} \; 100 \le t < 105 , \\
		1 & {\rm for} \; 105 \le t < 140 ,\\
		\frac{1}{2}-\frac{1}{2}\cos{\left[ \pi \frac{(145-t)}{5} \right]} & {\rm for} \; 140 \le t < 145 ,
		\end{array} \right. 
		\label{e7}
\end{equation}
with $t$ in units of simulation hours and represents a smooth ramp-up period followed by a uniform driving period and a smooth ramp-down back to zero for $t \ge 145$~hr. The function $f_B(B_r)$ defines the spatial extent over the surface by smoothly enforcing the range of radial field magnitudes over which to calculate the flow profiles to $|B_r(R_\odot, \theta, \phi)| \in [1,8]$~G as
\begin{equation}
f_B(B_r) = \left \{ \begin{array}{ll}
    		\sin{\left[ 2\pi \frac{(B_r+8)}{7} \right]} \;\; & {\rm for} \; -8.0 \le B_r \le -1.0 ,\\
		\sin{\left[ 2\pi \frac{(B_r-1)}{7} \right]} & {\rm for} \; +1.0 \le B_r \le +8.0 .
		\end{array} \right.   
\end{equation}
The coefficient $V_0 = \pm 4 \times 10^{15}$~cm$^2$~s$^{-1}$~G$^{-1}$ yields a maximum magnitude $\boldsymbol{V}_{\rm shear} \approx 20$~km~s$^{-1}$. This maximum flow speed ensures that the evolution during the energization phase is quasi-static, i.e. much less than both the Alfv\'{e}n speed in the vicinity of the global PIL, $V_{\rm shear}/V_A \lesssim 5$\%, and the sound speed, $V_{\rm shear}/c_0 \approx 10$\%.

Figure~\ref{fsFLOW}(a) plots the magnitude of the surface velocity components, $(V_\theta^2 + V_\phi^2)^{1/2}$, on the $r=R_\odot$ lower boundary at $t=110$~hr during the uniform shearing phase. The $\boldsymbol{V}_{\rm shear}$ distribution traces the global PIL underneath the stellar streamer belt. The small velocity magnitudes (dark blue) in the polar regions are the non-radial components of the steady-state stellar-wind outflow. Figure~\ref{fsFLOW}(b) plots representative magnetic field lines at $t=135$~hr, late in the shearing phase. The large-scale, sheared-arcade field structure above the global PIL is a common feature of extended filament channels on the Sun \citep{MacKay2010, Pevtsov2012}. Our field lines develop a weak twist from the structure of the boundary flows and form the characteristic dips found in many prominence field models and observations \citep{DeVore2000b, Parenti2014}. The distribution of mass density is shown in the plane of the sky to highlight the closed-flux-streamer belt region of the stellar corona.

\section{Simulation Results}
\label{sec:cme}

%
\begin{figure}[!h]
\centerline{ \includegraphics[width=0.47\textwidth]{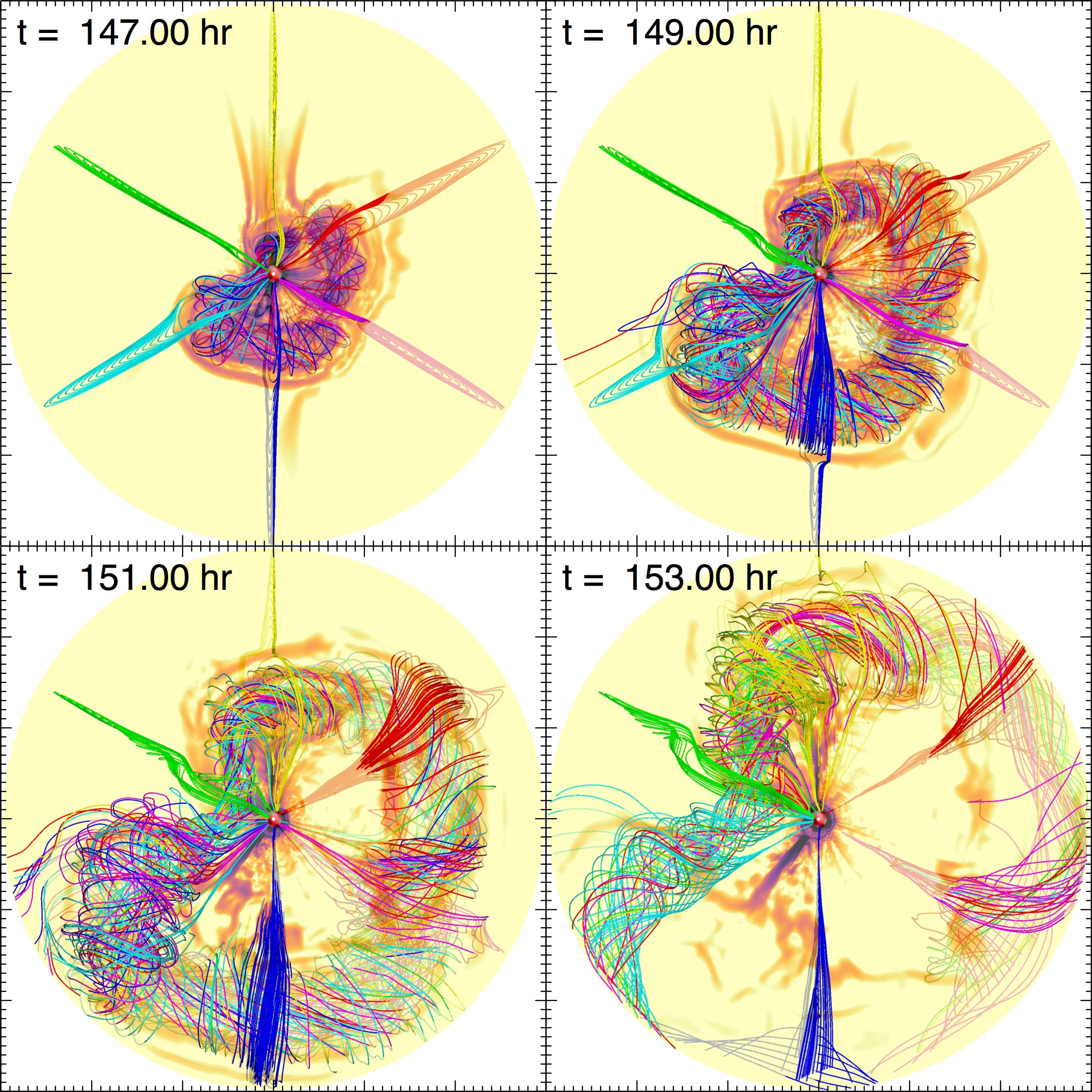} }
\caption{Evolution of the global CME flux rope propagating through the equatorial plane, viewed from the north stellar pole.\\ 
(An animation of this figure is available.)}
\label{f3}
\end{figure}

%
\begin{figure*}
\centering
\includegraphics[width=1.0\textwidth]{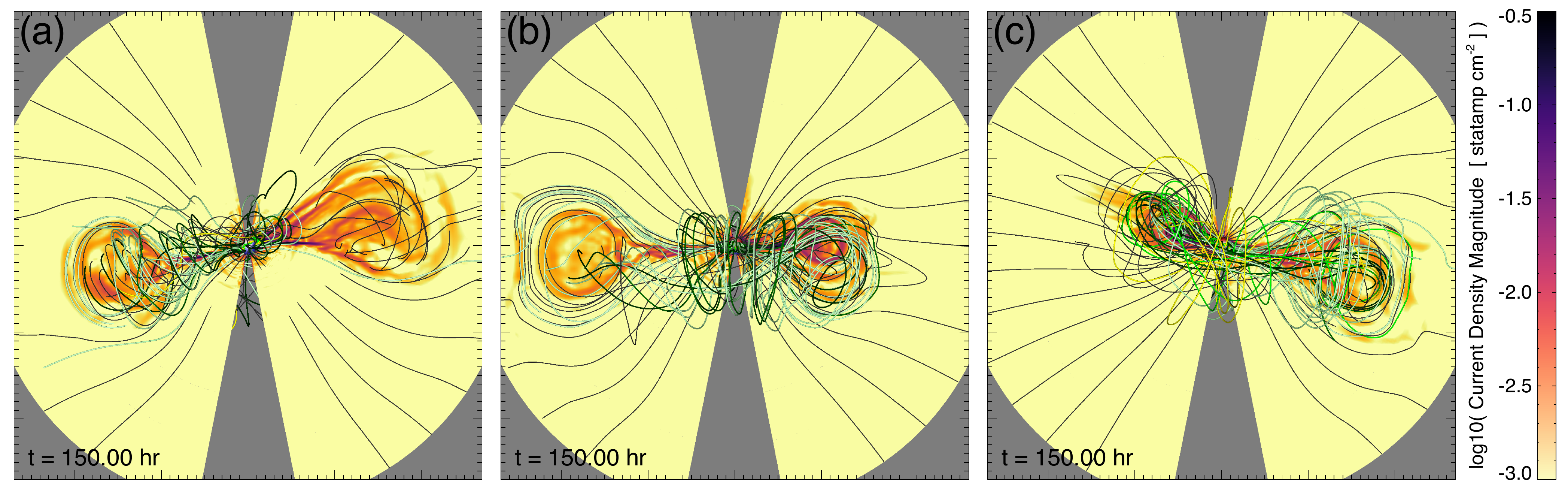}
\caption{Formation and eruption of the stellar CME flux rope structure at $t=150$~hr for central meridians: (a) $\phi = +120^{\circ}$; (b) $\phi=0^{\circ}$; (c) $\phi=-120^{\circ}$.\\
(An animation of this figure is available.)\\}
\label{fsCME}
\end{figure*}

%
\subsection{Carrington-scale Eruptive Stellar Flare and CME}
\label{sec:cme:energy}

The eruption process in our simulation follows the CSHKP scenario described by \citet{Lynch2016b} and references therein. The energized field slowly rises due to the force imbalance set up by the increased magnetic pressure of the sheared-flux core. A radial current sheet forms underneath the expanding sheared flux. Eventually, fast magnetic reconnection sets in at the current sheet facilitating the rapid release of stored magnetic energy, ejecting a coherent twisted flux rope, and rebuilding the closed-flux system as the post-eruption flare arcade. Due to the global scale of our eruption, the post-eruption flare arcade becomes the closed-flux streamer belt encircling the entire star. The global eruption and evolution of the super-CME are shown in Figure~\ref{f3} and its animation. The 3D structure of the erupting stellar CME flux rope can also be visualized from different vantage points in the ecliptic plane corresponding to the viewpoint of white-light coronagraph observations. Figure~\ref{fsCME} shows a snapshot of our stellar CME flux rope at $t=150$~hr from three different meridional perspectives in the ecliptic plane: (a) $\phi=+120^{\circ}$, (b) $\phi=0^{\circ}$, and (c) $\phi=-120^{\circ}$. The plane-of-the-sky contour plots show the electric current density magnitude on a logarithmic scale to illustrate the CME cross sections. Representative field lines are chosen to illustrate the transition of the magnetic structure of the pre-eruption sheared flux into the erupting CME flux rope and its propagation through the outer stellar corona. Figure~\ref{fsCME} is included as an animation.

The typical three-part structure of CMEs in coronagraph observations---a bright leading circular front, a dark circular cavity, and a bright central or trailing core region---is one of the best proxy measures of the magnetic structure of flux-rope CMEs \citep{Vourlidas2013}. Figure~\ref{fsCME} shows this characteristic magnetic structure in the plane-of-the-sky cross-sections of our global CME eruption. The figure also shows the magnetic structure of the flux rope propagating towards the observer in the ecliptic plane---the configuration of most halo-CME eruptions that impact the Earth and cause significant geomagnetic responses \citep{ZhangJ2007}. Understanding this connection between the pre-eruption magnetic configuration of the CME source region and the CME's structure and evolution during the eruption and propagation through the heliosphere is of critical importance to terrestrial space-weather forecasting \citep{Palmerio2018} and will play an increasingly important role in characterizing exoplanetary space weather \citep{Airapetian2016a,CohenO2018}.

\begin{figure}
\centering
\includegraphics[width=0.47\textwidth]{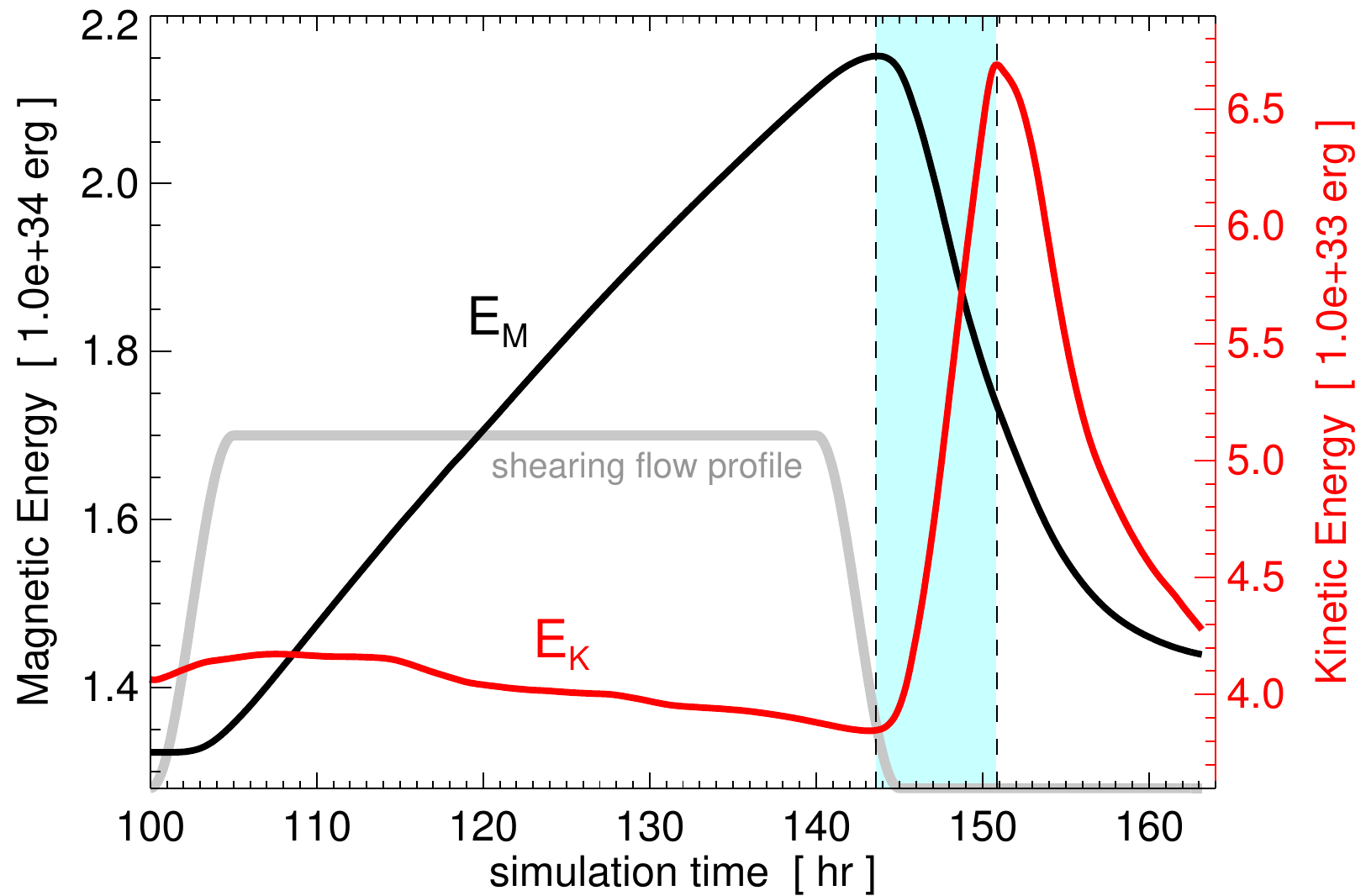}
\caption{Global magnetic energy ($E_M$, black) and kinetic energy ($E_K$, red) evolution during energization and eruption phases. Temporal profile of boundary shearing flows is shown in gray. Vertical dashed lines and light blue shaded region indicate the impulsive phase of the stellar eruptive flare.}
\label{f5}
\end{figure}

Figure~\ref{f5} plots the evolution of the global magnetic energy ($E_M$, black) and kinetic energy ($E_K$, red). Here the energization phase is indicated with the light gray line. $E_M$ increases as the applied shearing flows drive the accumulation of free magnetic energy. The light-blue shaded region, defined as $t_{\rm pre} \le t \le t_K$ where $t_{\rm pre} = 143.58$~hr is time of maximum $E_M$ and $t_K = 150.83$~hr is the time of maximum $E_K$, indicates the impulsive phase of the eruption where the energy conversion is most rapid. Defining $\Delta E_M(t) \equiv E_M(t_{\rm pre}) - E_M(t)$ and $\Delta E_K(t) \equiv E_K(t) - E_K(t_{\rm pre})$,
the total magnetic energy released by the end of the simulation, $t_f = 163.17$~hr, is $\Delta E_M(t_f) = 7.13 \times 10^{33}$~erg, the maximum increase in kinetic energy is $\Delta E_K(t_K) = 2.84 \times 10^{33}$~erg, and the magnetic-to-kinetic energy conversion ratio is $\Delta E_K(t_K) / \Delta E_M(t_K) = 68.5\%$ during the impulsive phase and $\Delta E_K(t_K) / \Delta E_M(t_f) = 39.9\%$ over the entire eruption process. The global scale of our eruption means that the stellar flare current sheet is reasonably well-resolved; consequently, our energy ratio is very similar to the $\approx$ 30\% found by \citet{Karpen2012} for a global eruption with adaptive mesh refinement. We note that the CME associated with the famous Carrington event of 1859 was estimated to have a kinetic energy $\Delta E_K \sim 2 \times 10^{33}$~erg \citep{Cliver2013}.

\subsection{Stellar Flare Reconnection Flux}
\label{sec:cme:rxn}

The flare ribbons and reconnection flux in the simulation data are calculated using a version of the  \citet{Kazachenko2017} methodology used to characterize two-ribbon flares in SDO data. However, instead of using SDO/AIA 1600\AA\ emission thresholding to determine whether a pixel is ``flaring,'' here we use the change in field line length $\Delta L \equiv L(t) - L(t-\Delta t)$ as a proxy for the rapid geometric re-configuration of the field-line connectivity between simulation data output intervals of $\Delta t = 20$~min. We create a $192 \times 384$ uniform grid in $(\theta, \phi)$ at the $r=R_\odot$ lower boundary and trace magnetic field lines from these footpoints for each simulation output time. If a long field line becomes significantly shorter by $\Delta L \le -3R_\odot$ over the $\Delta t$ interval between consecutive output files, then we consider that pixel to have undergone reconnection. The reconnection pixels are accumulated in time to create the cumulative ribbon area map.

\begin{figure}
\centering
\includegraphics[width=0.47\textwidth]{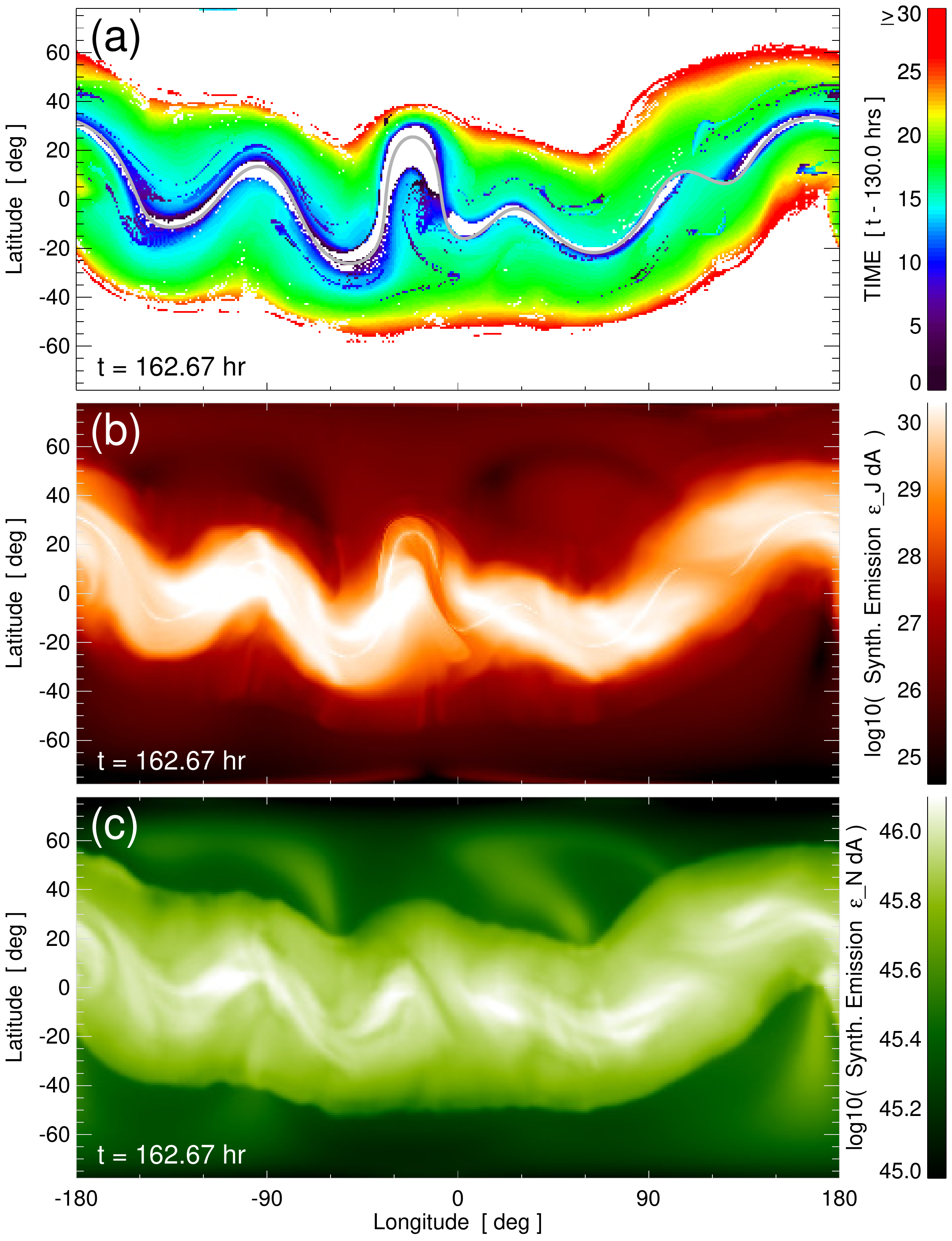}
\caption{Global eruptive flare ribbon structure and post-eruption arcade light-of-sight integrated emission distribution. Panel (a): Spatiotemporal distribution of the magnetic flux that reconnects during the eruptive stellar flare. Panel (b): Synthetic line-of-sight integrated hot (X-ray) coronal emission from the post-eruption flare arcade. Panel (c): Synthetic ambient (EUV) coronal emission in the same format as panel (b). \\
(An animation of this figure is available.)}
\label{fsFLARE}
\end{figure}

Figure~\ref{fsFLARE}(a) shows the time evolution of the area on the stellar surface swept out by the two-ribbon flare. The color scale indicates the first time the magnetic flux bundle at a given pixel has reconnected through the flare current sheet and become part of the post-eruption flare arcade. Large two-ribbon flares are often characterized by the ``zipper effect,'' where the ribbons form and rapidly grow parallel to the source region PIL and then move more slowly away from the PIL in the perpendicular direction \citep{Moore2001,Qiu2009,Linton2009,Aulanier2012,Priest2017}. This effect is clearly seen in the Figure~\ref{fsFLARE} animation.

From the time series of ribbon area pixel mask shown in Figure~\ref{fsFLARE}(a), we can calculate the stellar reconnection flux as
\begin{equation}
\Phi_{\rm rxn} = \sum_{j, k} |B_r(R_\odot,\theta_j, \phi_k)| \; dA_{jk} ,
\end{equation}
where the pixel area is the usual $dA_{jk} = R_\odot^2 \sin{\theta_j} \Delta \theta \Delta \phi$. The reconnection rate is then calculated as the central-differenced time derivative.

\begin{figure*}[t]
\centerline{ \includegraphics[width=1.0\textwidth]{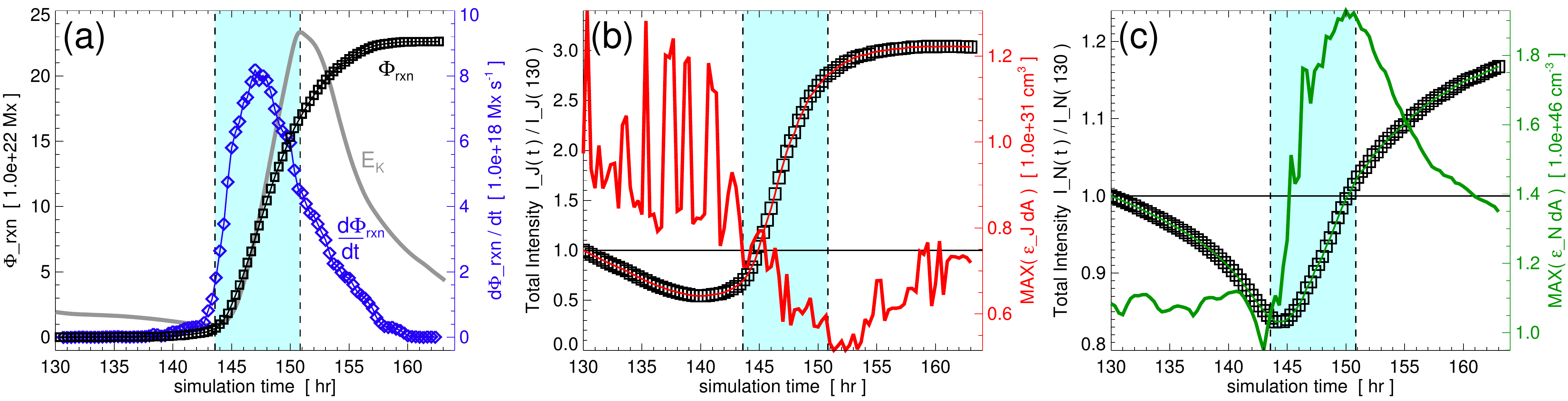} }
\caption{Reconnection flux and synthetic flare emission light curves. 
({a}) unsigned reconnection flux and reconnection rate. ({b}) Area-integrated light curve of mean hot (X-ray) intensity (squares) and maximum emissivity (red solid line). ({c}) Total light curve of ambient (EUV) intensity (squares) and maximum emissivity (green solid line).\\}
\label{f6}
\end{figure*}

Figure~\ref{f6}(a) shows the unsigned reconnection flux $\Phi_{\rm rxn}$ (black squares) and the reconnection rate $\dot \Phi_{\rm rxn}$ (blue diamonds) versus time, along with the global kinetic energy ($E_K$) shown in gray. The total unsigned reconnection flux at the end of the simulation ($t_f = 162.67$~hr) is $\Phi_{\rm rxn}(t_f) = 2.26 \times 10^{23}$~Mx, and the maximum reconnection rate is $\dot \Phi_{\rm rxn} = 8.0 \times 10^{18}$~Mx~s$^{-1}$.
The impulsive phase of the eruptive flare, defined previously, is highlighted by the light-blue shading. The stellar CME is driven by the eruptive flare reconnection, as seen by the substantial $\dot \Phi_{\rm rxn}$ increase \emph{before} the global kinetic energy increase.

%

\begin{figure}
\centering
\includegraphics[width=0.47\textwidth]{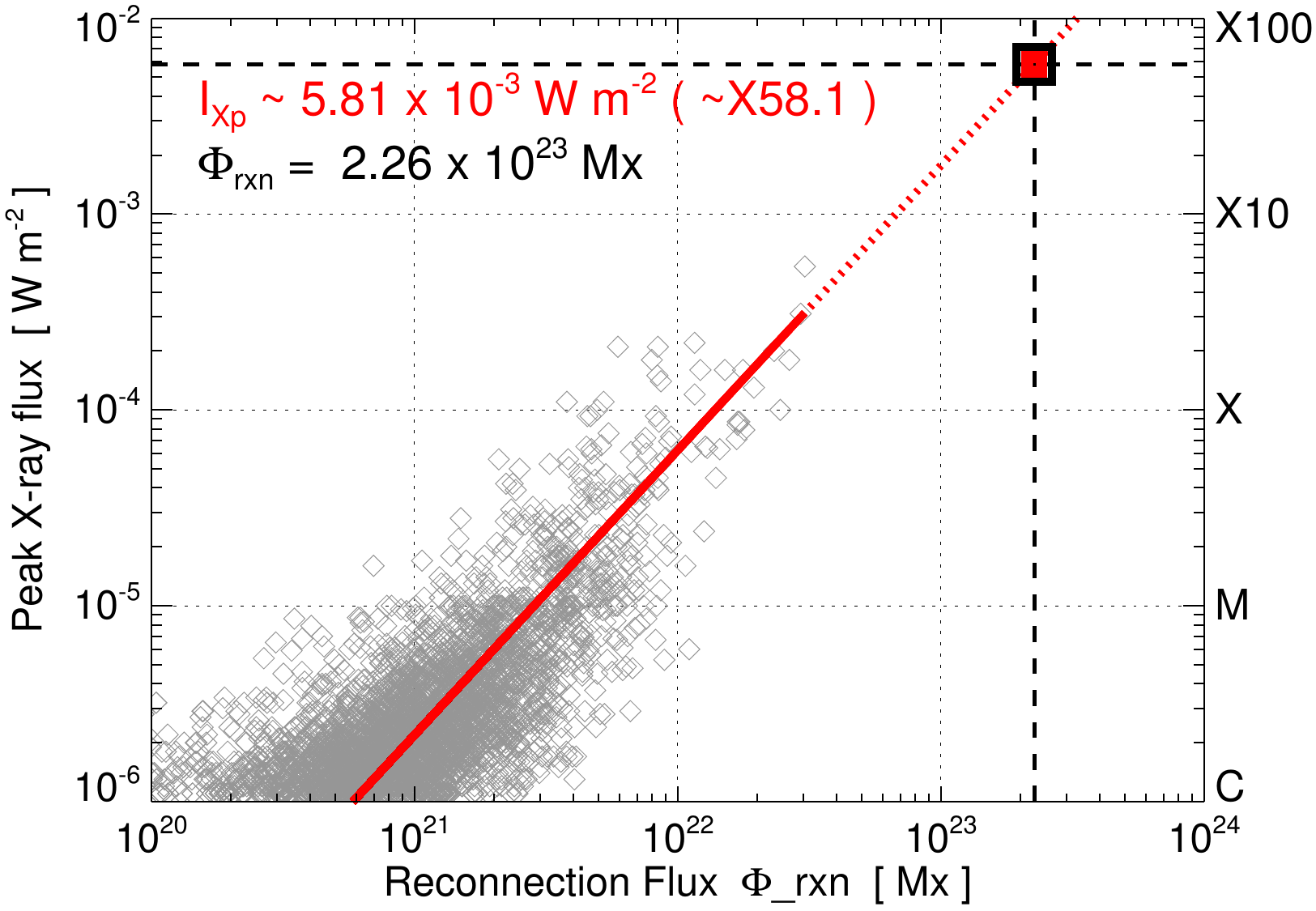}
\caption{Extrapolation of the simulation reconnection flux to an estimated X-ray flux based on the power-law fit obtained by the \citet{Kazachenko2017} analysis of two-ribbon flares in SDO/AIA and SDO/HMI data.}
\label{fs3}
\end{figure}

The \citet{Kazachenko2017} power-law relationship between GOES X-ray flux and reconnection flux was determined from 3137 solar flares of classes $\ge$C1.0 analyzed by in the SDO observations between 2010--2016. It is given by
$I_{\rm Xray} = \alpha \left( \Phi_{\rm rxn}/10^{21} \right)^{1.454}$, where $\alpha = 2.19 \times 10^{-6}$~W~m$^{-2}$~Mx$^{-1.454}$.
Figure~\ref{fs3} plots each of the \citet{Kazachenko2017} $(I_{\rm Xray}, \Phi_{\rm rxn})$ data points (gray diamonds) as well as the power-law fit (solid red line) for the range of their observations. The power law is continued (dotted red line) beyond the solar observations, and we have shown the location of our simulation reconnection flux on it as the red square.
The calculated X-ray flux of $I_{\rm Xray} = 5.8 \times 10^{-3}$~W~m$^{-2}$ corresponds to a X58 class flare comparable to the estimate for the 1859 Carrington event of an X45($\pm$5) class flare \citep{Cliver2013}.

%
%

%
\subsection{Synthetic Flare X-Ray and EUV Emission}
\label{sec:cme:flare}

To compare with general morphology and evolution of solar flare observations, we construct two synthetic emission proxies representing hot (10 MK) and ambient (1 MK) emission that we will refer to as synthetic X-ray and EUV emission, respectively.

The first emission proxy uses the method developed by \citet{Cheung2012}. The synthetic differential emissivity proxy $d\varepsilon_J$ at a three-dimensional point in space $\boldsymbol{r}_{ijk}$ is taken to be proportional to the average squared current density over the magnetic field line $\langle J^2 \rangle$, normalized by the total field-line length, $L$. The total synthetic emission intensity $\varepsilon_J$ is then the line-of-sight integral over this differential emissivity proxy. For simplicity, here we calculate the (spherical) differential emissivity cube as a $100 \times 192 \times 384$ grid in $(r, \theta, \phi)$ and take the line-of-sight to be in the radial direction for $r \in \left[ 1R_\odot, 2R_\odot \right]$. The synthetic emission is thus
\begin{equation}
\varepsilon_J(\theta,\phi) = \sum_i d\varepsilon_J( \boldsymbol{r}_{ijk} ) = \sum_i \langle J^2 \rangle_{ijk}
\end{equation}
where
\begin{equation}
\langle J^2 \rangle_{ijk} = \frac{1}{L} \int_0^{L} d\ell' \; | J( \boldsymbol{r}(\ell') ) |^2 
\end{equation}
for a magnetic field line that passes through $\boldsymbol{r}_{ijk}$ and is parameterized by a differential arc length $d\ell'$. Figure~\ref{fsFLARE}(b) plots the logarithm of line-of-sight-integrated emission $\varepsilon_J$ at $t=162$~hr corresponding to the spatial extent of the stellar flare arcade, which has become the entire closed-flux region of the stellar corona after the eruption. We note that $\varepsilon_J$ is qualitatively most similar to the SDO/AIA 131\AA\ emission where the filter bandpass captures lines of Fe~VIII associated with transition region temperature of $\log{T} \sim 5.6$ but also contains a significant flaring-corona contribution from Fe~XXI at $\log{T} \sim 7.0$ \citep{Lemen2012}. Therefore, we refer to our synthetic emission distribution and light curves based on $\varepsilon_J$ as synthetic X-ray emission, representing hot (10 MK) plasma temperatures.

Our second synthetic emission proxy, $d\varepsilon_N$ is taken as proportional to $n_e^2$ which we equate with the single-fluid plasma number density $n^2$ at point $\boldsymbol{r}_{ijk}$. This yields
\begin{equation}
\varepsilon_N(\theta,\phi) = \sum_i d\varepsilon_N( \boldsymbol{r}_{ijk} ) = \sum_i n^2(r_i,\theta_j,\phi_k) .
\end{equation}
Figure~\ref{fsFLARE}(c) plots the logarithim of the line-of-sight-integrated emission $\varepsilon_N$. Since $\varepsilon_N$ is qualitatively similar to SDO/AIA 171\AA\ emission, which is primarily an Fe~IX contribution from upper transition-region/quiet or ambient coronal temperatures of $\log{T} \sim 5.8$, our light curves based on $\varepsilon_N$ represent ambient (1 MK) emission and which we refer to as synthetic EUV emission.

The Figure~\ref{fsFLARE} animation also shows the spatiotemporal evolution of our synthetic flare emission proxies. We note that by $t=162$~hr, the global flare arcade encompasses the entire pre-eruption closed-flux region of the stellar corona.

To compare with general morphology and evolution of stellar flare light curve observations, we integrate the Figure~\ref{fsFLARE}(b) and \ref{fsFLARE}(c) spatial emission distributions. The surface-integrated light curves derived from our synthetic X-ray and EUV proxies are calculated as
\begin{equation}
I_{\{J,N\}} = \sum_{j,k} \varepsilon_{\{J,N\}}(\theta_j,\phi_k) \; dA_{j,k} \; ,
\end{equation}
and normalized by the pre-eruption values at $t=130$~hr. The total surface-averaged intensity light curves ($I_J$, $I_N$) are plotted in Figures~\ref{f6}(b) and \ref{f6}(c) as black squares, respectively. The maximum (radial) line-of-sight-integrated emission curves in Figure~\ref{f6}(b) and \ref{f6}(c) are taken as $\max\left[ \varepsilon_J dA \right]$ and $\max\left[ \varepsilon_N dA \right]$, respectively, and are plotted as solid lines (X-ray, red; EUV, green).

Our light curves all show a qualitative transition during the impulsive phase of the global eruptive flare, but each curve has a slightly different character. The X-ray mean intensity shows a $\approx$ 30\% pre-eruption dimming and a $\approx$ 200\% post-eruption brightening---reminiscent of the $\Phi_{\rm rxn}$; the maximum X-ray emissivity shows a highly variable pre-eruption enhancement transitioning to a less-variable post-eruption dimming of $\approx$ 40\%. In contract, the EUV mean intensity shows a pre-eruption dimming of $\approx$ 15\% that transitions during the flare to a modest enhancement of $\approx$ 18\% afterwards; and the maximum EUV emissivity shows the sharpest rise just after the onset of the flare, an $\approx$ 80\% enhancement, and the clearest post-eruption decay.  We note that, while solar and stellar X-ray and EUV observations measure the combined temperature and density evolution during flares, the synthetic emission from our isothermal simulation is determined solely by the density component.

Stellar flare observations in X-ray, UV, optical, and radio wavelengths are generally consistent with the CSHKP understanding of eruptive solar flares \citep{Hawley1995,Guedel2002,Osten2015}. \citet{Harra2016} discussed the solar SDO/EVE disk-integrated light curves of Fe ion spectral lines for signatures that could be applied to stellar flare observations. They found the lower ionization states of Fe show prompt dimming curves, whereas the higher ionization states show a rapid increase followed by a slow decay. The stellar observations of pre-flare dimming have competing physical interpretations \citep{Leitzinger2014,Osten2017}.  The origin of the pre-flare dimming in our mean intensity light curves can be understood from the spatiotemporal evolution of the emission distributions shown in the Figure~\ref{fsFLARE} animation. As the pre-eruption closed-flux regions expand and the outer layers open into the stellar wind, the emission from the streamer belt decreases gradually. This is followed by the rapid CME-related dimming due to the eruption/opening-up of most of the remaining closed flux once the flare reconnection begins in earnest. The eruption-related dimming is coincident with the initial formation of the flare arcade that has significantly enhanced emission compared to the pre-flare configuration. As the post-eruption flare arcade grows, its enhanced emission becomes the dominant feature of the global mean intensity light curves.

\begin{figure*}[t]
\centerline{ \includegraphics[width=1.0\textwidth]{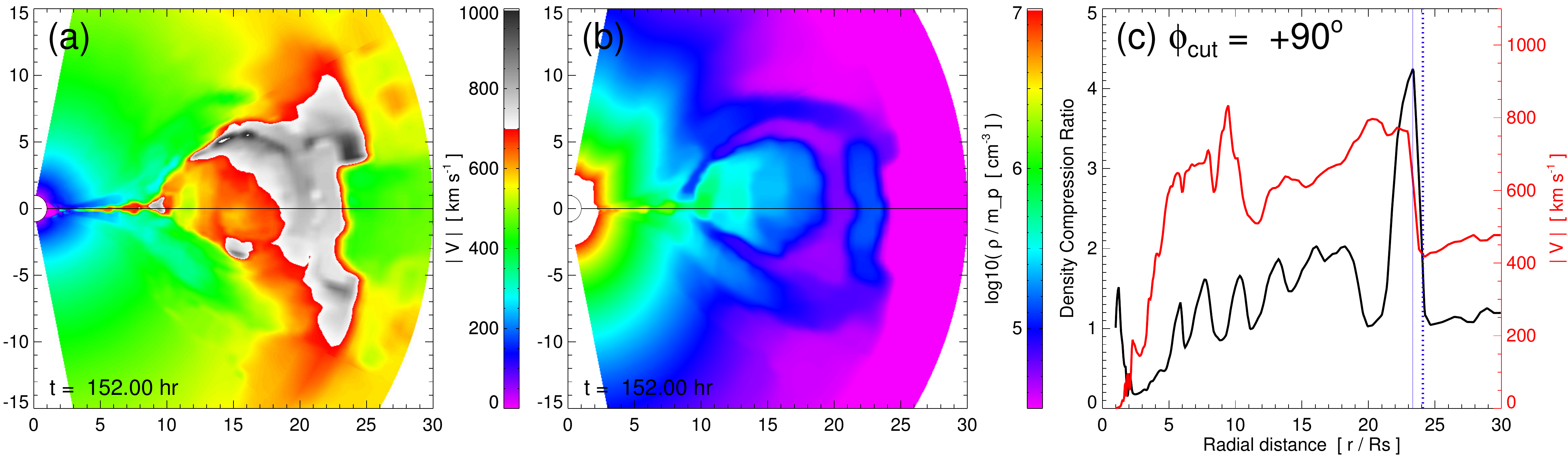} }
\caption{Radial cuts through CME density and velocity magnitude at $\phi_{\rm cut}=+90^{\circ}$. ({a}) Velocity magnitude. ({b}) Number density. ({c}) compression ratio and $|V|$ along radial cut. }
\label{f7}
\end{figure*}


\subsection{Stellar CME and CME-driven Shock Parameters}
\label{sec:cme:shock}

CMEs are responsible for some of the most geoeffective space-weather impacts at Earth and other solar system bodies. The combination of southward-directed $B_z$ in the sheath and ejecta flux rope with the increased dynamic pressure in dense sheath regions driven by fast events can cause significant geomagnetic responses \citep{ZhangJ2007,Li2018}. Fast CMEs drive shocks, and these coronal and interplanetary shocks are often sites of substantial energetic particle acceleration \citep{Lario2016,Luhmann2018}. Our simulation results can be used to begin an investigation of stellar CMEs and CME-driven shocks, as a precursor to estimating exoplanetary space-weather.

Figures~\ref{f7}(a) and \ref{f7}(b) show the distributions of velocity magnitude and number density, respectively, in the meridional plane at $\phi_{\rm cut}=+90^{\circ}$ during the global eruption. The black lines indicate the radial sampling we use to quantify the CME-driven shock properties. Figure~\ref{f7}(c) shows the shock density compression ratio (black) and velocity magnitude (red) for the $\phi_{\rm cut} = +90^{\circ}$ meridional plane. The density compression ratio is calculated as $d = n(r,t) / n(r,130)$ so that the upstream, unperturbed values $\approx 1$. The shock location is determined along the radial sampling trajectory by calculating the radial gradient of the compression ratio $\partial d(r,t)/\partial r$ and choosing the maximum (negative) value at the largest radial distance---shown as the vertical blue dotted line. The peak number density of the stellar CME sheath region is found as the maximum number density within a spatial window just downstream of the shock location (shown as the solid blue line).

CME-driven shock strengths can be estimated from white-light coronagraph observations.  \citet{Ontiveros2009} examined a number of shock fronts driven by fast ($\ge$1500~km~s$^{-1}$) CMEs and showed their compression ratios ranged from 1.5--3 in the LASCO C3 field of view. \citet{Kwon2018} used multi-viewpoint STEREO observations of two fast halo CMEs to show how the compression ratios were strongest at the nose of the shock and weaker at the flanks. \citet{Manchester2008b} presented a comparison of the CME-driven shock in their numerical simulations of the $\approx$ 2000~km~s$^{-1}$ halo CME of 28 Oct 2003 with the coronagraph observations, showing good agreement with the observed compression ratio of $\approx$ 5 at $r \sim 15R_\odot$. In Figure~\ref{f7}(c), our compression ratio reaches $\approx$ 4 by $25R_\odot$, indicative of a strong shock.

\begin{figure*}
\centerline{ \includegraphics[width=1.0\textwidth]{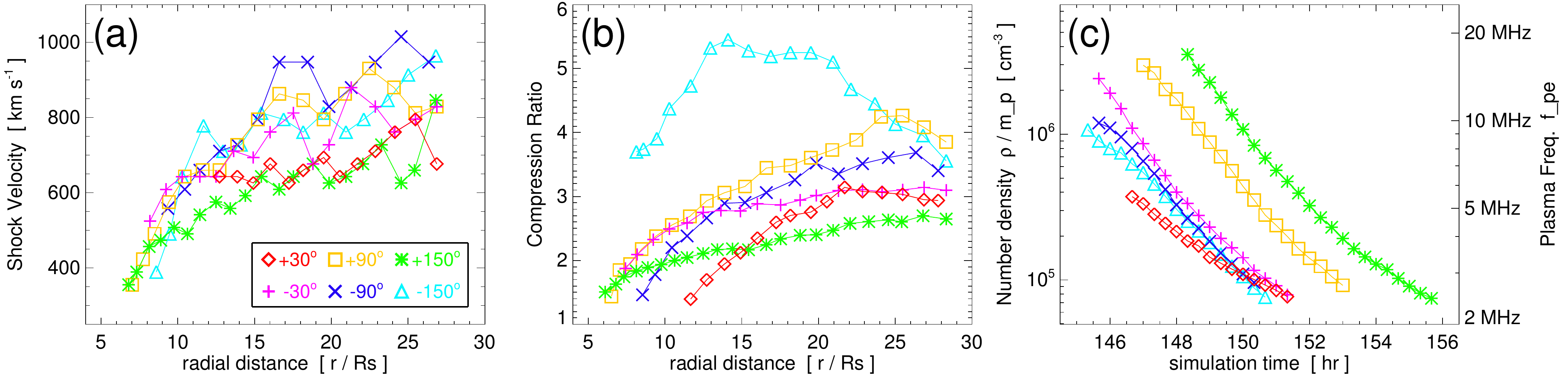} }
\caption{CME-driven shock properties along six radial cuts: 
({a}) radial velocity; 
({b}) compression ratio; 
({c}) density and plasma frequency.
}
\label{f8}
\end{figure*}

Figure~\ref{f8} shows the spatiotemporal evolution of physical properties at the location of the peak compression ratio of the CME-driven shock for the six radial cuts through our simulation at longitudes $\phi_{\rm cut} \in \{ \pm30^{\circ}, \pm90^{\circ}, \pm150^{\circ} \}$. The latitude of each radial cut was chosen to approximate the center of the erupting CME flux rope cross-section in each meridional plane. 
{
Figure~\ref{f8}(a) shows the shock velocities, \ref{f8}(b) the shock compression ratios, and \ref{f8}(c) the shock densities and plasma frequencies. In (b) at some locations, the compression ratio significantly exceeds 4, which is the upper bound for a plasma with adiabatic index $\gamma = 5/3$. This can occur in our isothermal model, for which $\gamma = 1$ and the compression ratio is theoretically unlimited, as well as in the more general polytropic models, for which $1 < \gamma \ll 5/3$ and the compression ratio has a large but finite upper bound. Irrespective of the energy model that is used, large compression ratios reliably indicate the occurrence of strong shocks with Mach numbers $M \gg 1$---the key requirement for shock acceleration of energetic particles.
}

The shock strength (compression ratio), the magnetic field strength, and the shock orientation angle with respect to the magnetic field are the main inputs into models for the shock-acceleration of solar or stellar energetic particles \citep[SEPs; e.g.][]{Zank2007}. A greater shock strength and $B$ magnitude result in a higher energy roll-over of the spectrum (a ``harder'' spectrum) that increases the fluxes of the highest energy particles, e.g., protons $\gtrsim50$~MeV. The ambient particles that are swept up by the shock and then accelerated are referred to as the seed particle population. The upstream density and turbulence determine the number of seed particles and energizes their distribution, respectively. The downstream turbulence/waves generated in quasi-parallel shocks and low cross-field diffusion at quasi-perpendicular shocks both contribute to greater SEP energization.

The higher density and velocities in the ambient solar and stellar winds of the young Sun and analogs like $\kappa^{1} Cet$, can drive shocks in the stream interaction regions, providing greater and more highly energized seed particle populations \citep{Airapetian2016c} that will enhance SEP production by CME-driven shocks. Stellar energetic particle fluxes may significantly impact exoplanetary atmospheres. For example, \citet{Airapetian2016b} modeled the atmospheric chemistry of the early Earth under large SEP event conditions, showing that precipitating energetic protons enhanced chemical reactions that convert molecular nitrogen, carbon dioxide, and methane into nitrogen oxide and hydrogen cyanide.

Solar flares and CMEs often generate radio emission signatures. For example, CME-driven shocks can produce Type II radio bursts \citep{Gopalswamy2005,Schmidt2016} during their propagation through the corona and interplanetary space and flare-accelerated electron beams often produce a variety of Type III bursts \citep{Reid2014,McCauley2018}.

To estimate the radio emission from electrons at our CME-driven shocks, we use the standard formulation \citep{Huba2013} where the electron plasma frequency $f_{pe}$ depends on $n_e$ $\left[ {\rm cm}^{-3} \right]$ as
\begin{equation}
f_{pe} = 8.98 \times 10^3 \; n_e^{1/2} \; \left[ {\rm Hz} \right] .
\end{equation}
Figure~\ref{f8}(c) shows the number density and corresponding plasma frequency and we obtain a range of characteristic frequency drifts of $\Delta f/\Delta t \approx$ 0.6--2.0~MHz~hr$^{-1}$ with initial plasma frequency emission in the 5--20~MHz range. Given the apparent difficulty in observing radio signatures of stellar CMEs thus far \citep[e.g.][]{Crosley2018, Mullan2019}, the use of modeling to constrain the parameter space of stellar wind and CME densities and velocities that should result in observable signatures will be increasingly important. We note that this is a zeroth-order estimate of radio emission frequencies associated with our MHD fluid properties. In general, in order to estimate synthetic radio fluxes that can be directly compared to observations, a number of additional factors are required including source geometry considerations, signal propagation effects, and the specific details of various non-thermal emission mechanisms, as discussed by \citet{Moschou2018} and references therein.

%
\section{Discussion}
\label{sec:disc}

%
%
%

Our simulation self-consistently models both gradual accumulation and rapid release of free magnetic energy during the eruption of a global-scale stellar CME from the measured background magnetic field of $\kappa^{1}Cet$. The total magnetic flux and energy contained in the equatorial streamer belt of a star provides a baseline estimate for the anticipated strengths of its eruptive superflares/superCMEs. The energization of the background field of $\kappa^{1}Cet$ and the ensuing impulsive phase of stellar flare reconnection in our simulation created and accelerated a global CME eruption with kinetic energy $\Delta E_K \approx 3\times10^{33}$~erg and duration $\Delta t_K \approx$ 10~hr. This energy is similar to the estimated energies of both the extreme 1859 Carrington event from the Sun \citep{Cliver2013} and a 2016 naked-eye superflare from the nearby M-dwarf star Proxima Centauri \citep{HowardW2018}. Our results show that even the comparatively weak background magnetic fields of such stars, relative to the estimated strengths of the fields in starspots, can store sufficient energy to power detectable superflares. We found that the free energy that could be stored in our pre-eruptive configuration for $\kappa^{1}Cet$ reached $E_M(t_{\rm pre}) / E_M(t_{\rm rel}) - 1 \approx 63\%$. This is very close to the free energy ($66\%$) required to open to infinity all of the magnetic field lines of an elementary dipole configuration \citep{Mikic1994}. The amount of energy that was converted to kinetic energy of our eruption is $\Delta E_K(t_K) / E_M(t_{\rm rel}) \approx 22\%$, or approximately one-third of the stored free energy.

The energetic superflare on $\kappa^{1}Cet$ observed by \citet{Robinson1987} was estimated by \citet{Schaefer2000} to have released about $2\times10^{34}$ erg, an order of magnitude more energy than our simulated event from the recently measured background magnetic field of the star \citep{Rosen2016}. It is interesting to note that the energy of the observed 1986 $\kappa^{1}Cet$ superflare is roughly twice the energy contained within its background field, $2\times10^{34}$ erg vs.\ $1\times10^{34}$ erg (cf. Figure~\ref{fs2}). Correspondingly, the estimated energy of the historic Carrington 1859 solar flare \citep{Cliver2013} and the energy of the Sun's present-day background field \citep{Yeates2018} are each smaller by about one order of magnitude, $2\times10^{33}$ erg vs.\ $1\times10^{33}$ erg. 
{The similarity in the flare-to-background energy ratio suggests that the Robinson-Bopp superflare on $\kappa^{1}Cet$ may have been as extreme an event for that star as the Carrington flare was for the Sun. Verifying this correspondence would require obtaining and analyzing a long-duration database of flare observations of $\kappa^{1}Cet$ and similar stars. Encouragingly, 
\citet{Notsu2019} and collaborators are investigating precisely this subset of superflaring solar-type stars in the \textit{Kepler} data. 
}

%
%

{
There are two primary areas where our idealized modeling could be improved upon in future numerical simulations. First, the magnetic flux distribution in the ZDI stellar magnetogram captures the global structure of the background magnetic field but does not resolve the field strengths or spatial scales associated with starspots or active regions. Second, the isothermal model for the stellar atmosphere does not account for the full thermodynamic evolution of the plasma, which affects the density and temperature distribution of the background wind and also means that the flare-related energy deposition into bulk plasma heating during the eruption is not captured in the simulation. We discuss each of these issues, in turn, below.
}

%
%
{
Our energization process imparts shear and twist to the coronal field structure in exactly the place the Sun requires (i.e., localized above the polarity inversion line). This enables the system both to gradually accumulate free energy and to transition rapidly to an unstable, runaway eruption that removes this localized stressed field from the closed-field corona via magnetic reconnection. Thus, the energization and eruption-triggering \emph{processes} in our simulation are completely generic and should be universally applicable, including to more realistic solar and stellar magnetic field configurations.  
}

{
More realistic stellar active region fields will be stronger (100s--1000s~G) and significantly more localized spatially, although potentially over areas larger than observed on the Sun. For example, \citet{Rucinski2004} used the \textit{Microvariability and Oscillations of Stars} (MOST) observations of $\kappa^{1}Cet$ to derive the sizes of two large starspots which covered 1.4\% and 3.6\% of the stellar surface, respectively. These areas are roughly 10 times greater that the largest sunspot/AR areas ever observed on the Sun \citep{Hoge1947}. The magnetic energy estimates used in stellar flare analyses are typically of the form $E_M \sim (1/(8\pi)) B_{AR}^{2} A_{AR}^{3/2}$ \citep[e.g.][]{Shibata2013, Maehara2015, Notsu2019}. If we were to scale the magnetic energy of our system, $E_M(t_{rel}) \sim 1.3 \times 10^{34}$~erg, to a starspot area estimate of $A_{AR} \sim f (\pi R_{\star}^{2})$ where $f = 0.002-0.02$, then we obtain a range of areas $A_{AR} = 3 \times 10^{19-20}$~cm$^2$ and a resulting range of starspot/AR field strengths, $B_{AR} \sim 250-1400$~G. These field strengths are comfortably within the range of both solar observations \citep[e.g. Figure 10 of][]{Kazachenko2017} and the \citet{Saar1992} estimate of $350-500$~G for the $\kappa^{1}Cet$ surface-averaged field magnitudes. 
}

{
We emphasize that our approach represents an attempt to model the most extreme stellar space-weather event possible within the observational constraints imposed by the surface magnetic-flux distribution from the ZDI reconstruction. Despite the global spatial scale and the stellar magnetogram's unresolved starspot/AR flux distribution, the magnetic energy stored and released in our simulation's eruptive superflare is compatible with the order-of-magnitude estimate one obtains from typical starspot/AR areas and magnetic field strengths. On the other hand, if much stronger unresolved starspots/AR flux distributions were assumed, the resulting superflare and super-CME energies could be much higher. For example, recent analyses by \citet{Okamoto2018} of \emph{Hinode} spectropolarimetry data of NOAA AR 11967 in 2014 Feb 1--6 showed a peak magnetic field strength of 6.2~kG. This solar magnetic field magnitude over the range of starspot/AR areas above would result in much larger magnetic energy estimates of $E_M \sim 2.5 \times 10^{35} - 8 \times 10^{36}$~erg. 
}

%
%

{
Our use of an isothermal stellar atmosphere neglects the internal energy equation in the MHD system and consequently, the evolution of our gas pressure is determined solely by the variation in mass density. In more complex thermodynamic MHD models, the internal energy equation typically includes field-aligned heat conduction, radiative cooling, magnetic and viscous dissipation, and a source term for the local contribution from coronal heating. One such parameterized coronal heating model represents the dissipation of energy associated with Alfv\'{e}nic turbulence, which contributes both to local plasma heating and, through the wave pressure gradient, to the acceleration of the solar wind \citep{vanderHolst2014, Lionello2014, Oran2017}. While a polytropic stellar wind outflow naturally creates a significant mass density gradient between the open-flux regions (coronal holes) and the closed-field corona (the helmet streamer belt), a more realistic treatment of the internal energy equation is expected to increase this density contrast for both the quiet-Sun and active regions \citep[e.g.][]{Torok2018}. The isothermal model for our stellar atmosphere has two main consequences for our simulation results: first, on the structure and intensities associated with our synthetic EUV and soft X-ray emission profiles; second, on the interaction between the CME and the background wind, including the properties of the CME-driven shock. 
}

{
As discussed above, approximately one-third of the free magnetic energy stored prior to the eruption is converted to kinetic energy of the CME. About one-sixth of the stored energy remained in the closed coronal field after the eruption; that is, the magnetic field did not relax completely to the initial, minimum-energy state of the system. The remaining energy released by the magnetic field during the eruption, $\Delta E_H / E_M(t_{\rm rel}) \approx 32\%$ of the initial magnetic energy, was not captured by our simple isothermal model. This remnant released energy $\Delta E_H$ would appear as thermal energy due to magnetic and viscous dissipation, heating the stellar plasma to high flare temperatures and being radiated away into space. With a more realistic temperature structure in the dynamic formation and evolution of the flare current sheet and the post-eruption arcade \citep[e.g.][]{Reeves2010, Lynch2011}, one could improve the synthetic emission calculations of $\S$\ref{sec:cme:flare} by calculating the density- and temperature-dependent emission intensities in various spectral lines and convolving these with instrument response functions \citep[e.g.][]{Lionello2009, Reeves2010, Shen2013b, Oran2017, Jin2017b}. 
}

{
Relaxing the isothermal assumption would result in a modified stellar wind profile and, thus, quantitative differences in the plasma properties of the CME--stellar wind interaction region. Qualitatively, the overall picture will remain the same as in Figure~\ref{f7}, i.e., the energetic eruption driven by the flare reconnection will create and accelerate a rapidly expanding, highly-magnetized (low $\beta$) flux rope ejecta which will generate a shock and a dense compression region at the CME's leading edge. A modified upstream stellar wind profile will obviously impact the absolute number densities and an improved thermodynamic treatment will allow for additional localized, compressional heating. The shock is expected to remain Alfv\'{e}nic, however the detailed shock parameters such as the compression ratio, field strength, and the shock normal orientation will have different quantitative values. Even with the simple isothermal model used here, the compression ratios generated in our simulation are broadly consistent with those determined from the white-light coronagraph observations of large solar events.        
}

{Therefore, limitations notwithstanding,} our simulation results represent an important step toward understanding observations of superflares from active solar-type stars and characterizing CMEs from magnetically active stars across the K--M dwarf-star spectrum. We analyzed the properties of CME-induced shocks, including the frequency and duration of associated Type II events that may be detected in future low-frequency (10 MHz or lower) radio observations of magnetically active stars. The derived properties of CMEs and their associated shocks can provide inputs to models of stellar SEP energization and transport via the diffusive shock acceleration mechanism \citep{LiGang2013,HuJ2018}. Knowledge of the SEP energy spectrum and particle fluence is critical for evaluating biogenic conditions on terrestrial-type exoplanets around active stars, as well as for the early Earth and Mars. \citet{Airapetian2016a} have shown that large SEP fluxes can increase the production rates of nitrous oxide, a powerful greenhouse gas, and hydrogen cyanide, a feedstock molecule for prebiotic synthesis. Characterizing the SEP environment will help to determine the boundaries of the planetary ``biogenic'' zone \citep{Airapetian2019} and will be important in specifying the efficiency of ozone destruction and surface dosages of ionizing radiation that are damaging to life on planetary surfaces. Future modeling efforts that describe the conditions of CME initiation in {more realistic stellar magnetic field distributions} can help to understand observations of G, K, and M dwarf stars using the currently implemented international multi-observatory program (TESS, Hubble, XMM-Newton, Apache Point Observatory) and to prepare for next-generation JWST observations.

\acknowledgments

BJL, MDK, and WPA acknowledge support from NASA NNX17AI28G, NSF AGS-1622495, and  the Coronal Global Evolutionary Model (CGEM) project NSF AGS-1321474. VSA acknowledges support from the NASA Exobiology 80NSSC17K0463 and the TESS 1 Cycle 1 program. CRD acknowledges support from NASA H-SR and LWS TR\&T programs. TL acknowledges support via the Austrian Space Application Programme (ASAP) of the Austrian Research Promotion Agency (FFG) within ASAP11 and the FWF NFN project S11601-N16. OK acknowledges support by the Knut and Alice Wallenberg Foundation (project grant ``The New Milky Way''), the Swedish Research Council (project 621-2014-5720), and the Swedish National Space Board (projects 185/14, 137/17). The computational resources for this work were provided to BJL by the NASA High-End Computing Program through the NASA Center for Climate Simulation at Goddard Space Flight Center.

%



\bibliographystyle{apj} 
\bibliography{/Users/blynch/bibliography/master}

\end{document}